\documentclass{article}
\usepackage{graphicx}

\usepackage[colorlinks, citecolor=blue, linkcolor=blue, backref=true]{hyperref}
\usepackage{longtable}
\usepackage{url}
\usepackage{amsthm}

\theoremstyle{definition}

\theoremstyle{remark}

\usepackage[firstinits=true, sorting=none,parentracker=true, doi=true, isbn=false, natbib=true]{biblatex}

\bibliography{./biblio}
\usepackage{longtable}
\usepackage[toc,page]{appendix}
\usepackage{amsmath}
\usepackage{amssymb}
\usepackage{eufrak}
\usepackage{epstopdf}
\usepackage{color}

\begin{document}
\title{Scattering on square lattice from crack with damage zone}
\author{Basant Lal Sharma\thanks{Department of Mechanical Engineering, Indian Institute of Technology Kanpur, Kanpur, U. P. 208016, India ({bls@iitk.ac.in})}
\and
Gennady Mishuris\thanks{Department of Mathematics, Aberystwyth University, Wales, UK
({ggm@aber.ac.uk})
}}
\maketitle

\begin{abstract}
{A semi-infinite crack in infinite square lattice }is subjected to a wave coming from infinity, thereby leading to its scattering by the crack surfaces.
{A partially damaged zone ahead of the crack-tip is modeled} by an arbitrarily distributed stiffness of the damaged links.
While the open crack, with an atomically sharp crack-tip, in the lattice has been solved in closed form with help of {the} scalar Wiener-Hopf formulation
(SIAM Journal on Applied Mathematics, 75, 1171--1192; 1915--1940), the problem considered {here}
becomes very intricate depending on the nature of damaged links.
For instance, in the case of partially bridged finite zone it involves a $2\times2$ matrix kernel of formidable class.
But using an original technique, the problem, including the general case of arbitrarily damaged links, is reduced to a scalar one with the exception that it involves solving an auxiliary linear system of $N \times N$ equations where $N$ defines the length of the damage zone.
The proposed method does allow, effectively, the construction of an exact solution.
Numerical examples and the asymptotic approximation of the scattered field far away from the crack-tip are also presented.
\end{abstract}

\section*{Introduction}

Among other distinguished as well as popular works \cite{chadwick2012continuum}, Peter Chadwick has made several contributions to the wave propagation problems in anisotropic models with different kinds of symmetries as well as those applicable to theory of lattice defects \cite{chadwick1964application,chadwick1970wave,chadwick1977foundations,chadwick1982surface,chadwick1989wave,chadwick1987basic,chadwick1997application}.
The researches on elastic cubic crystals are specially relevant in the context of the present paper as a discrete counterpart of square lattice is natural when one considers waves interacting with a crack-tip \cite{slepyan1,slepyananti,slepyan2002models,sharma2015diffraction,sharma2015diffraction3,sharma2016guides}.

\begin{figure}[!ht]
\center{\includegraphics[width=.4\linewidth]{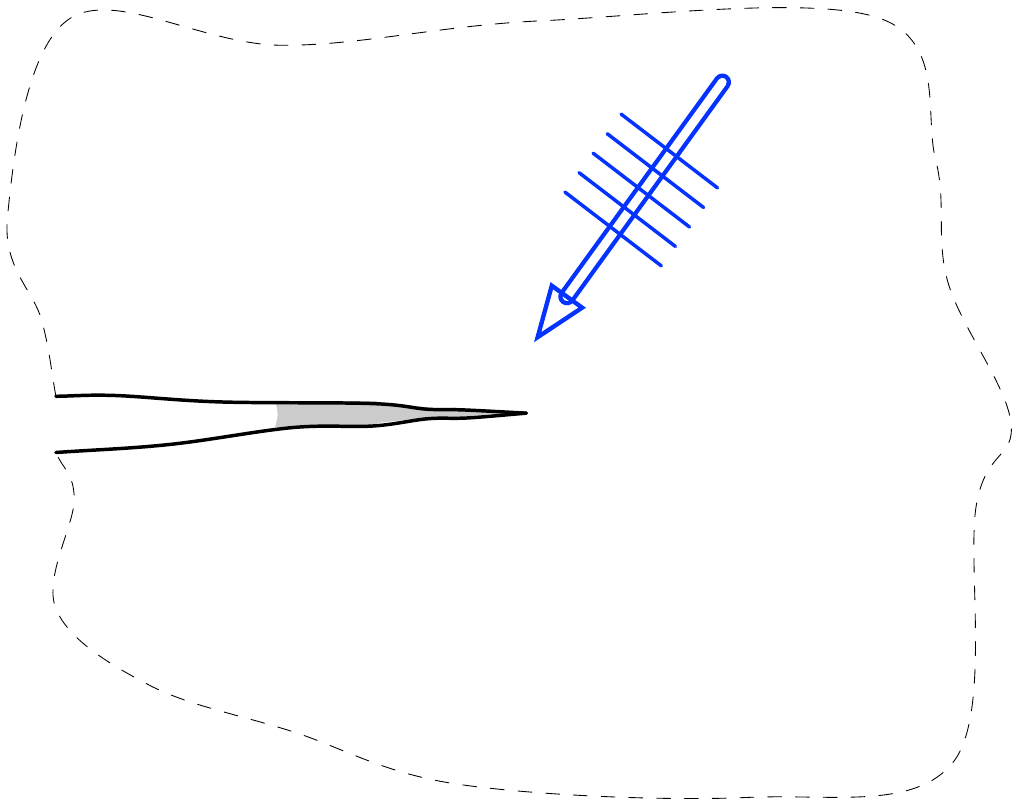}}
\caption[ ]{Schematic of the incident wave on a crack-tip with damage}
\label{fig0}
\end{figure}

Indeed, the role of discrete models in the description of mechanics and physics of crystals \cite{BornHuang1985} and related structures has been dominant in studies of several critical phenomena {like dislocation dynamics, dynamic fracture and phase transition, bridge crack effects, resonant primitive, localised and dissipative waves in lattices among others} \cite{atkinson1965,celli1970,thomson1971lattice,slepyanI,Kulak,marder1,marder,Slepyan_F1,Slepyan_F2,Slepyan_F3,Movchan_2007,mishuris2007waves,MMS_2008,ASS_2008,MMS_2009,Slepyan_2010,sharma_balance}. The concomitant issues dealing with the propagation of waves interacting with {stationary} cracks and rigid {constraints} {as well as surface defects have} been explored in  \cite{CNJMM_2013,sharma2015diffraction,sharma2015diffraction3,sharma2016guides,Gaurav2019_asymp,Gaurav2019_pair,sharma2017scattering,sharma2015diffraction2,sharma2015diffraction4,sharma2016diffractionedge,sharma2015diffraction5,MMJC_2017,Sharma2019_surface}.
{It is noteworthy that the continuum limit, that is low frequency approximation, of the scattering problem for a single crack \cite{sharma2015diffraction3,Blslimit}, recovers the well known solution of Sommerfeld \cite{Sommerfeld,Sommerfeld1}.
With respect to the crack-tip geometry, note that the discrete scattering problems have been solved in  \cite{sharma2015diffraction,sharma2015diffraction3,sharma2015diffraction4,sharma2015diffraction5,Gaurav2019_pair} for atomically sharp crack tips.
}
Typically, such situations of discrete scattering due to crack surfaces are further complicated as the crack-tip is endowed with some structure, as schematically shown in Fig. \ref{fig0}, due to the presence of
cohesive zone, partial bridging of bonds, etc, {commonly used in continuum mechanics
\cite{dugdale1960yielding,barenblatt1962mathematical,cox1994concepts}.
The notion of cohesive zone used in this paper is considered in a wider sense than in the fracture mechanics (it is not clearly eliminating any singularities that are not arising in discrete formulation). The zone simply emphasizes the fact that different links, subjected to a high amplitude vibration, near the crack-tip may undergo phase transition, damage and/or even breakage at different time depending on materials properties (manifesting by respective damage/fracture criteria \cite{GVVM_2018}).
As a result, a naturally created partial bridging and/or forerunning zone can be observed during the crack propagation
(see for example \cite{MMS_2009,NMS_2017}).}

The problem considered in this paper, in fact, becomes much more intractable when compared with the scattering due to an atomically sharp crack-tip that has been solved in \cite{sharma2015diffraction,sharma2015diffraction3}
using the scalar Wiener-Hopf factorization \cite{noble,slepyan2002models}.
As an example, it is shown that in the case of partially bridged finite zone, the corresponding Wiener-Hopf problem becomes vectorial as it involves a $2\times2$ matrix kernel {which} belongs to a formidable class \cite{abPhase,abPlatesI,abPlatesII,abPlatesIII}.
In this paper, it is shown that a reduction to a scalar problem is possible with the additional clause that it involves solving an auxiliary linear system of $N \times N$ equations where $N$ represents the size of the cohesive zone.
Such reduction resembles {the one} proposed for the Wiener--Hopf kernel with exponential phase factors in continuum case \cite{abPhase,abPlatesI,abPlatesII,abPlatesIII}, and its recently investigated discrete analogue of scattering due to a pair of staggered crack tips \cite{Sharmastaggerpair_exact,Gaurav2019_asymp}.
It is also relevant to recall for such kernels an asymptotic factorization based alternative, {but} approximate, approach \cite{mishuIII,Gaurav2019_asymp}.

Overall, the method proposed in the paper does allow, effectively, the construction of an exact solution, even in the general case of arbitrary set of damaged links.
The paper presents some numerical examples to demonstrate the effect of certain {kind} of damaged links on the pattern of scattered field. The expression obtained after an asymptotic approximation of the scattered field far away from the crack-tip is also presented as a perturbation over and above that for {the} atomically sharp crack-tip obtained earlier in \cite{sharma2015diffraction}.
A careful analysis of the continuum limit \cite{Blslimit}, in the presence of damaged links which demands adoption of a proper scaling, is relegated to future work. The question of the behavior of edge conditions vis-a-via  sharp {cracks} \cite{Silver,noble} is anticipated to be crucial in such exercise.

As a summary of the organization and presentation of the main aspects of this paper, \S\ref{problemform} gives the mathematical formulation of the scattering problem.
\S\ref{solutionWH} provides the exact solution of the Wiener--Hopf equation modulo the reduced form to an auxiliary linear system of $N \times N$ equations.
\S\ref{specialdamage} presents some special scenarios of the distribution of the damaged links which allow either {an} immediate solution of the auxiliary equation or demonstrate the difficulty and richness of the problem by mapping its difficulty to a class of problems.
\S\ref{farfield} gives the far-field behaviour away from the crack-tip as a perturbation in addition to that for a sharp crack-tip, as well as some numerical examples.
\S\ref{concl} concludes the findings of this paper.
One appendix appears in the end giving technical details of application of the Wiener--Hopf method.
For {details of the theory of} scattering and {the} Wiener-Hopf method we refer to
\cite{morse,noble} whereas
the mathematical aspects of convolution integrals and Fourier analysis can be found in
\cite{Paley,Wiener,Mikhlin,gakhov,titI,KreinGoh,Bottcher}.
For the issues dealing with the difficult cases of {the} matrix Wiener-Hopf problems, the reader is referred to \cite{Speck1,Meistersys1,Meistersys2,daniele1984solution,mishuIII,mishuI,mishuII,MishuIV}

\section{Problem formulation}
\label{problemform}
Let us consider a square lattice structure consisting of a semi-infinite crack that involves an additional structural feature near the crack{-}tip.
The bulk lattice is constructed {with} the same masses, $m$, situated {at} the points $(x,y)$, $x{\in \mathbb{Z}},y\in \mathbb{Z}$ and connected by elastic springs with stiffness, $c>0$ (see Fig. \ref{fig0}).
The space{-}coordinates are dimensionless and define the position of the corresponding mass $(x,y)=(\tilde x/a,\tilde y/a{)}$ (normalised by the length of the links between the neighboring masses $a$). Displacement of the mass at each point is denoted as $u_{x,y}(t)$.

\begin{figure}[!ht]
\center{\includegraphics[width=.9\linewidth]{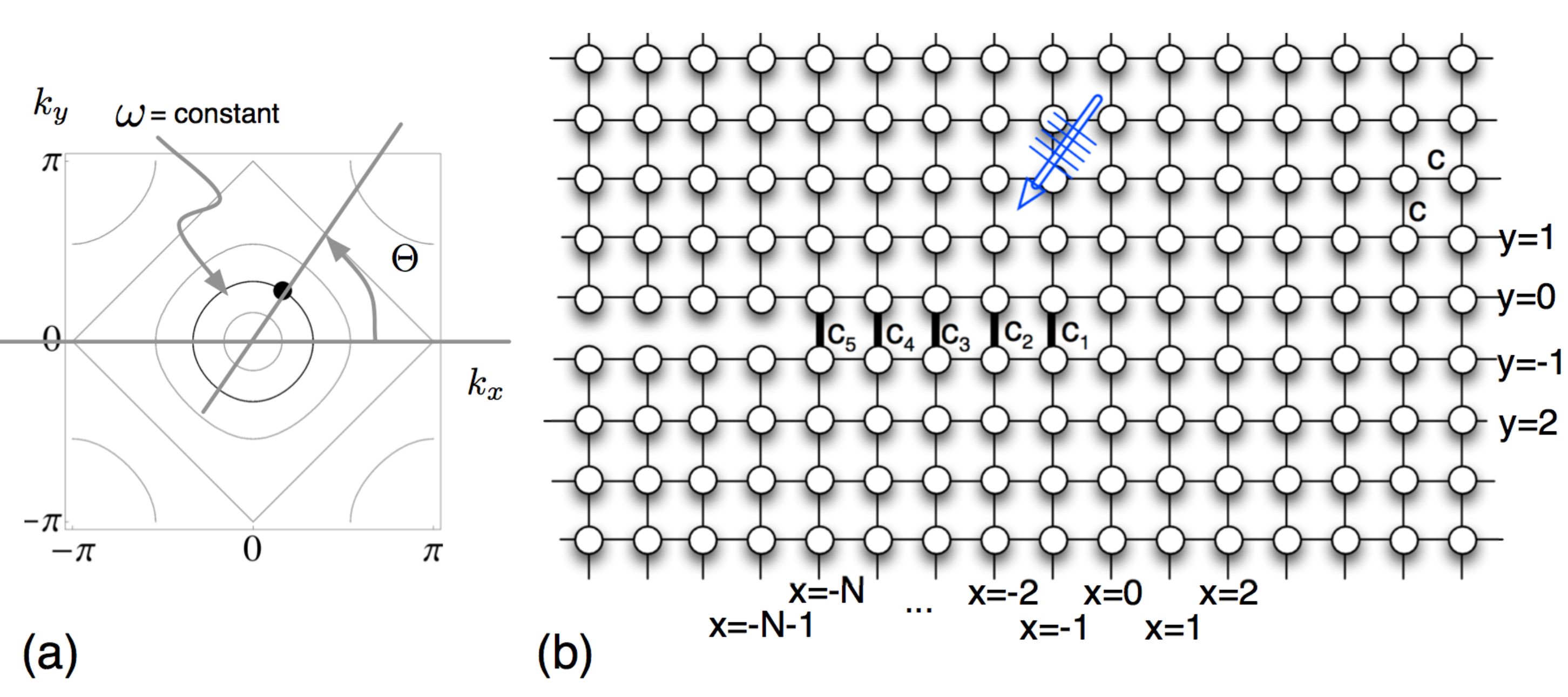}}
\caption[ ]{(a) Schematic of the incident wave parameters relative to the typical contours for square lattice dispersion relation. (b) Geometry of the square lattice structure and the notation for the number of damaged sites $N=5.$
}
\label{fig1}
\end{figure}

The bonded interface between the two half-planes consists of a finite segment of distributed springs of the stiffness, $\{c_{-x}\}_{x=-1}^{-N}$
{($c_{-x}\ge0$)} that connecting masses from the different sides of the interface attributed to the values of the variable $x$ (see Fig. \ref{fig1}).
{Note that some of the links can {be also considered fully} destroyed{;} thus{,} the geometry of the damage zone can be rather complex.}

In the following, we will use the standard notation:
\begin{equation}
\mathbb{Z}^+=\{0,1,2,\ldots\}, \quad \mathbb{Z}^-=\{-1,-2,\ldots\},\quad {\mathbb{Z}= \mathbb{Z}^+\cup {\mathbb{Z}^-}.}
\label{eq:different_Z}
\end{equation}

We assume that an incident wave
\begin{equation}
u^i(x,y,t)=
u^i_{x,y}e^{-i\omega t}=Ae^{-ik_xx-ik_yy-i\omega t},
\label{eq:incident}
\end{equation}
imposes the out-of-plane {small} deformation of the lattice. Here $k_x, k_y\in\mathbb{R}$ are wave numbers; also, sometimes we use $k_x=k\cos\Theta, k_y=k\sin\Theta$ with $k>0$ and $\Theta\in[-\pi, \pi]$. The symbol $A\in \mathbb{C}$ is {the} complex dimensional amplitude of the wave. It is further assumed that $\omega=\omega_1+i\omega_2$ (where $\omega_2>0$ is an arbitrary small number). The latter guarantees the causality principle to be addressed. Note that this implies $k=k_1+ik_2$ where $k_2$ is small when $\omega_2$ is small.
We {seek} the harmonic solution to the problem {of} the form:
\begin{equation}
u^t(x,y,t)=u^t_{x,y}e^{-i\omega t}=(u^s_{x,y}+u^i_{x,y})e^{-i\omega t},
\label{eq:full solution}
\end{equation}
where $u^s_{x,y}$ and $u^i_{x,y}$ are the scattered {part} and the incident part, {respectively}.

The following set of equations are valid in each part of the lattice structure outside the interphase ($y\ge1$, and $y\le-2$):
\begin{equation}
c\Delta u_{x,y}+\omega^2 u_{x,y}=0, \quad x\in \mathbb{Z}.
\label{eq:main_equation}
\end{equation}
Here $\Delta$ is the discrete {Laplace} operator {with $\Delta u_{x,y}=u_{x+1,y}+u_{x-1,y}+u_{x,y+1}+u_{x,y-1}-4u_{x,y}$} (see \cite{Collatz,slepyan2002models,sharma2015diffraction}) {and} in the following $u_{x,t}=u^t_{x,y}$.

The interphase consists of two lines $y=0$ and $y=-1$ (see Fig. 1).
Let the damaged portion be denoted by the values of the coordinate $x$ lying in
\begin{equation}
\mathcal{D}=\{-1, -2, \dotsc, -N\}.
\label{eq:S}
\end{equation}
Let us denote the Kronecker delta by {the symbol} $\delta${; it is} equal to unit
when $x\in \mathcal{D}$ and zero otherwise. Also we denote the discrete Heaviside function by $H(x)$ for $x\in\mathbb{Z}$, defined such that
$H(x)=1$, if $x\in\mathbb{Z}_+$, {while} $H(x)=0$ when $x\in \mathbb{Z}_-$.
Further, let us introduce the notation
\begin{equation}
{v_{x}=(u_{x,0}-u_{x,-1}),} \quad
v^{i,s}_{x}=(u^{i,s}_{x,0}-u^{i,s}_{x,-1}), \quad x\in \mathbb{Z}.
\label{eq:funk_v}
\end{equation}
As a result,
for $x\in \mathbb{Z}$,
the conditions, $c\Delta u_{x,0}+(c-c_{-x}\delta_{\mathcal{D},x}-cH(x))v_{x}+\omega^2 u_{x,0}=0,$ and $c\Delta u_{x,-1}-(c-c_{-x}\delta_{\mathcal{D},x}-cH(x))v_{x}+ \omega^2 u_{x,-1}=0$, linking the top part of the lattice with the bottom one, can be written as
\begin{equation}
c\Delta u^s_{x,0}+(c-c_{-x}\delta_{\mathcal{D},x}-cH(x))v^s_{x}+\omega^2 u^s_{x,0}=-(c-c_{-x}\delta_{\mathcal{D},x}-cH(x))v^i_{x},
\label{eq:sy=0b}
\end{equation}
\begin{equation}
c\Delta u^s_{x,-1}-(c-c_{-x}\delta_{\mathcal{D},x}-cH(x))v^s_{x}+ \omega^2 u^s_{x,-1}=(c-c_{-x}\delta_{\mathcal{D},x}-cH(x))v^i_{x}.
\label{eq:sy=-1b}
\end{equation}
The skew symmetry follows immediately, i.e.,
\begin{equation}
u^s_{x,-1}+u^s_{x,0}=0,\quad x\in \mathbb{Z},
\label{eq:skew}
\end{equation}
and in general $u^s_{x,-y-1}+u^s_{x,y}=0$, $y\in\mathbb{Z}^+$.
Hence, it is enough to look at $y=0$, or a difference of above equations \eqref{eq:sy=0b} and \eqref{eq:sy=-1b}.
Let $\mathcal{A}$ be an appropriate annulus in the complex plane, same as that stated in
\cite{sharma2015diffraction}, i.e.,
\begin{equation}
\mathcal{A}:=\{z\in\mathbb{C}\!: R_+<|z|<R_-\},\quad R_+=e^{-k_2}, \quad R_-=e^{k_2\cos\Theta}.
\label{eq:annul}
\end{equation}
Taking into account the skew symmetry of the problem
under consideration (see \cite{sharma2015diffraction} and \eqref{eq:skew}), we conclude that
\begin{equation}
v^F=2u_{0}^F.
\end{equation}

{Applying {the} Fourier transform
\begin{equation}
u^F(z)\equiv\sum_{x\in\mathbb{Z}}z^{-x}u^s_x,\quad
z\in\mathcal{A},
\label{fourier}
\end{equation}
to the equation \eqref{eq:main_equation} for scattering waves in the upper space $y\ge0$ we obtain following {Slepyan \cite{slepyan2002models} and Sharma \cite{sharma2015diffraction}}:
\begin{equation}
u_{y}^F(z)=\lambda^y(z) u_{0}^F(z),\quad y=0,1,2,\ldots,\quad
z\in\mathcal{A},
\label{fourier_up}
\end{equation}
with
\begin{equation}
\lambda(z)= \frac{r(z)-h(z)}{r(z)+h(z)}, \quad h(z)=\sqrt{Q(z)-2},\quad r(z)=\sqrt{Q(z)+2},
\label{lambda}
\end{equation}
and
\begin{equation}
Q(z)=4-z-z^{-1}-\omega^2.
\label{Q(z)}
\end{equation}
}

{An} analogous result can be obtained in the lower space ($y\le-1$). The details are identical to those for crack without damage zone as provided in \cite{sharma2015diffraction}.

Taking into account {the} condition \eqref{eq:skew} as well as \eqref{fourier_up} (in particular, $u^F_{1}=-u^F_{-2}=u^F_{0}\lambda$ with $\lambda$ given by \eqref{lambda}, we obtain
\begin{equation}
c{\big(\lambda-1-Q(z)\big)}v^F{(z)}+2c v_-{(z)}-c\mathcal{P}^s(z)=-2c v^i_-{(z)}+c\mathcal{P}^i(z).
\label{eq:sy=0d}
\end{equation}
Thus,
with $v^F=v_++v_-$,
we have\begin{equation}(\lambda-Q-1)(v_++v_-)+2 v_-=-2 v^i_-+\mathcal{P}^i+\mathcal{P}^s,\label{eq:sy=0f}\end{equation}
{i.e.,}
\begin{equation}v_++\left(\frac{2}{\lambda-Q-1}+1\right)v_-=-\frac{2}{\lambda-Q-1} v^i_-+\frac{1}{\lambda-Q-1}\Big(\mathcal{P}^i+\mathcal{P}^s\Big).\label{eq:sy=0g}\end{equation}Simplifying further
for $z\in\mathcal{A}$,
we get
\begin{equation}
v_++Lv_-=(1-L) v^i_--\frac{1}{2}(1-L)\big(\mathcal{P}^i(z)+\mathcal{P}^s(z)\big),
\label{eq:sy=0h}
\end{equation}
where $L(z)=h(z)/r(z)$,
while $\mathcal{P}^s$ (and $\mathcal{P}^i$)
is a polynomial in $z$ given by
\begin{equation}
\mathcal{P}^{s,i}(z)={\frac{2}{c}}\sum_{x\in\mathcal{D}}c_{-x}v^{s,i}_xz^{-x}.
\label{eq:sy=0e}
\end{equation}

{E}quation \eqref{eq:sy=0h} is {the} Wiener--Hopf equation for the Fourier transform of the bonds $v_\pm$ in the cracked row {($x\in \mathbb{Z}^\pm$)}. By inspection and comparison with the results for a single crack {without damage zone} obtained in \cite{sharma2015diffraction} reveals that the kernel remains the same but there is a presence of an extra unknown polynomial on the right hand side of the Wiener--Hopf equation.

\section{Solution of {the} Wiener--Hopf equation}
\label{solutionWH}
The relevant multiplicative factorization of the kernel $L$ in \eqref{eq:sy=0h}
on the annulus $\mathcal{A}$, i.e., $L=L_+L_-$ has been obtained
{in an explicit form in Equation (2.27) from \cite{sharma2015diffraction}.}
Thus, using this fact, \eqref{eq:sy=0h} can be written as
\begin{equation}
L_+^{-1}v_++L_-v_-=\mathtt{C}\quad  \text{ on }\quad \mathcal{A},
\end{equation}
where $\mathtt{C}=\mathtt{C}^a+\mathtt{C}^{\mathrm{P}^s}$ and
\begin{equation}
\begin{aligned}
\mathtt{C}^a(z)&=(L_+^{-1}(z)-L_-(z)) v^i_-(z), \\
\mathtt{C}^{\mathrm{P}}(z)&=-\frac{1}{2}(L_+^{-1}(z)-L_-(z))\big(\mathcal{P}^i(z)+\mathcal{P}^s(z)\big).
\end{aligned}
\end{equation}
Note that $L_+$ is analytic and non-vanishing for $|z|>R_+$ and that $L_-$ is analytic and non-vanishing for $|z|<R_-$.
Now
\begin{equation}
v^i_-(z)=A(1-e^{ik_y})\delta_{D-}(zz_{\mathtt{P}}^{-1}),\quad z_{\mathtt{P}}=e^{-ik_x},
\label{vimin}
\end{equation}
(notice that $|z_{\mathtt{P}}|>R_-$ in \eqref{eq:annul}), so that
\begin{equation}
\mathtt{C}^a=\mathtt{C}^a_++\mathtt{C}^a_-,
\label{eqnCa}
\end{equation}
with \cite{sharma2015diffraction}
\begin{equation}
\mathtt{C}^a_+=\big(L_+^{-1}-L_+^{-1}(z_{\mathtt{P}})\big)v^i_-,\quad \mathtt{C}^a_-=\big(-L_-+L_+^{-1}(z_{\mathtt{P}})\big)v^i_-.
\end{equation}
{Here} $\mathtt{C}^a_+$ is analytic for $|z|>R_+$ and that $\mathtt{C}^a_-$ is analytic for $|z|<R_-$.
Let $\mathcal{P}^t(z)$ denote the sum $\mathcal{P}^s(z)+\mathcal{P}^i(z)$ ({with} the coefficients $v^{t}_{x}=v^{s}_{x}+v^{i}_{x}$), i.e.,
\begin{equation}
\mathcal{P}^t(z)=\frac{2}{c}\sum_{x\in\mathcal{D}}c_{-x}v^t_xz^{-x}.
\label{defPt}
\end{equation}
Further (recall that {$\mathcal{D}$ is defined in \eqref{eq:S}})
\begin{equation}
L_+^{-1}(z)\mathcal{P}^t(z)=\frac{2}{c}\sum_{x\in\mathcal{D}}c_{-x}v^t_x\left(L_+^{-1}(z)z^{-x}\right), \quad z\in\mathcal{A}.
\label{defLplusPt}
\end{equation}
Using the expressions from \cite{sharma2015diffraction3}, $L_+^{-1}$ can be expanded in a series of the form
\[
L_+^{-1}(z)=\sum_{m\in\mathbb{Z}^+}\overline{l}_{+m}z^{-m},
\]
for $|z|>R_+$.
Thus, (with $x=-\nu\in\mathbb{Z}^-$)
\begin{equation}
L_+^{-1}(z)z^{-x}=\sum_{m\in\mathbb{Z}^+}\overline{l}_{+m}z^{-m}z^{\nu}
=\phi^x_+(z)+\phi^x_-(z), \quad z\in\mathcal{A},
\label{sum_1}
\end{equation}
where
\begin{equation}
\phi^x_+(z)=\sum_{m=\nu}^\infty\overline{l}_{+m}z^{-m}z^{\nu},\quad
\phi^x_-(z)=\sum_{m=0}^{\nu-1}\overline{l}_{+m}z^{-m}z^{\nu},
\label{sum_1_plus}
\end{equation}
and the first term is analytic outside a circle of radius $R_+$ while the second is analytic inside a circle of radius $R_-$ in the complex plane.
Therefore, in the context of \eqref{defLplusPt},
\begin{equation}
L_+^{-1}\mathcal{P}^t=\frac{{2}}{c}\sum_{x\in\mathcal{D}}c_{-x}v^t_x\phi^x_++\frac{{2}}{c}\sum_{x\in\mathcal{D}}c_{-x}v^t_x\phi^x_-.
\end{equation}
Above {additive} splitting {of $L_+^{-1}\mathcal{P}^t$}, naturally, allows the following additive decomposition,
\begin{equation}
\mathtt{C}^{\mathrm{P}}=\mathtt{C}^{\mathrm{P}}_++\mathtt{C}^{\mathrm{P}}_- \quad \text{ on }\quad \mathcal{A},
\label{eqnCP}
\end{equation}
where
\begin{equation}
\mathtt{C}^{\mathrm{P}}_+=-\frac{1}{c}\sum_{x\in\mathcal{D}}c_{-x}v^t_x\phi^x_+,\quad
\mathtt{C}^{\mathrm{P}}_-=-\frac{1}{c}\sum_{x\in\mathcal{D}}c_{-x}v^t_x\phi^x_-+\frac{1}{2}L_-\mathcal{P}^t,
\end{equation}
which are analytic outside and inside of circle of radius $R_+$ and $R_-$ in the complex plane, respectively.
{
As a final step, following the analysis in \cite{sharma2015diffraction} and using the expressions \eqref{eqnCa},  \eqref{eqnCP} and \eqref{eq:sy=0h} leads to
\begin{equation}
\begin{aligned}
L_+^{-1}(z)v_+(z)=\mathtt{C}^a_+(z)+\mathtt{C}^{\mathrm{P}}_+(z)+\chi(z), \quad |z|>R_+,\\
L_-(z)v_-(z)=\mathtt{C}^a_-(z)+\mathtt{C}^{\mathrm{P}}_-(z)-\chi(z), \quad |z|<R_-,
\end{aligned}
\end{equation}
where $\chi$ is an arbitrary polynomial in $z$ and $z^{-1}$. It can be shown in \cite{sharma2015diffraction} that as $z\to\infty$,
$$
L_+^{-1}v_+-\mathtt{C}^a_+-\mathtt{C}^{\mathrm{P}}_+\to \text{constant},
$$
while as $z\to0$,
$$
L_-v_--\mathtt{C}^a_--\mathtt{C}^{\mathrm{P}}_-\to0,
$$
so that, as a consequence of Liouville's theorem, $\chi\equiv0$.
Hence,
\begin{equation}
\begin{aligned}
v_+(z)=L_+(z)(\mathtt{C}^a_+(z)+\mathtt{C}^{\mathrm{P}}_+(z)), \quad |z|>R_+,\\
v_-(z)=L_-^{-1}(z)(\mathtt{C}^a_-(z)+\mathtt{C}^{\mathrm{P}}_-(z)), \quad |z|<R_-.
\label{vpmsol}
\end{aligned}
\end{equation}
Due to \eqref{vpmsol}
\begin{equation}
\begin{aligned}
v^F(z)=L_+(z)(\mathtt{C}^a_+(z)+\mathtt{C}^{\mathrm{P}}_+(z))+L_-^{-1}(z)(\mathtt{C}^a_-(z)+\mathtt{C}^{\mathrm{P}}_-(z))\\
=v^F_a(z)+v_{\mathrm{P}}^F(z),\quad z\in\mathcal{A},
\label{invfourier_up2}
\end{aligned}
\end{equation}
with
\begin{equation}
v^F_a=L_+\mathtt{C}^a_++L_-^{-1}\mathtt{C}^a_-,\quad
v_{\mathrm{P}}^F=L_+\mathtt{C}^{\mathrm{P}}_++L_-^{-1}\mathtt{C}^{\mathrm{P}}_-.
\end{equation}
Also the total field $v$ {(the total oscillatory filed along the symmetry axis)} is given by
\begin{equation}
\begin{aligned}
v_{x}&=v^i_{x}+v^s_{x}=v^i_{x}+\frac{1}{2\pi i}\int_{\mathtt{C}}v^F(z)z^{x-1}dz,\quad x\in\mathbb{Z}.
\label{invfourier_v}
\end{aligned}
\end{equation}
}
In particular, {expanding \eqref{vpmsol}${}_2$ further},
\begin{equation}
v_-=\left(-1+L_-^{-1}L_+^{-1}(z_{\mathtt{P}})\right)v^i_--\frac{1}{c}L_-^{-1}\sum_{x\in\mathcal{D}}c_{-x}v^t_x\phi^x_-+\frac{1}{2}\mathcal{P}^t,
\label{vmsol}
\end{equation}
with $|z|<R_-$. Re-arranging \eqref{vmsol}, we get
\begin{equation}
v_-(z)+v^i_-(z)=L_+^{-1}(z_{\mathtt{P}})\dfrac{v^i_-(z)}{L_-(z)}-\frac{1}{c}L_-^{-1}(z)\sum_{x\in\mathcal{D}}c_{-x}v^t_x\phi^x_-(z)+\frac{1}{c}\sum_{x\in\mathcal{D}}c_{-x}v^t_xz^{-x},
\label{vmsol1}
\end{equation}
Let $\mathfrak{P}_\mathcal{D}$ denote the projection of Fourier coefficients of a typical $f_-(z)$ for $|z|<R_-$ to the set $\mathcal{D}$, then above equation \eqref{vmsol1} leads to
\begin{equation}
\sum_{x\in\mathcal{D}}\left(1-\frac{c_{-x}}{c}\right)v^t_xz^{-x}+\sum_{x\in\mathcal{D}}\frac{c_{-x}}{c}v^t_x\mathfrak{P}_\mathcal{D}\big(\frac{\phi^x_-}{L_-}\big)(z)
=\frac{\mathfrak{P}_\mathcal{D}\big(\dfrac{v^i_-}{L_-}\big)(z)}{L_+(z_{\mathtt{P}})}, \quad |z|<R_-,
\label{eqnNbyN}
\end{equation}
which yields a $N\times N$ system of linear algebraic equations for $\{v^t_x\}_{\mathcal{D}}$, i.e.,
the unknowns $\{v^s_x\}_{\mathcal{D}}$, since $\{v^i_x\}_{\mathcal{D}}$ are known in terms of the incident wave \eqref{eq:incident}.
Indeed, with the notation $\mathfrak{C}_{\kappa}(p)$ to denote the coefficient of $z^\kappa$ for polynomials $p$ of the form $\mathfrak{C}_1z+\mathfrak{C}_2z^2+\dotsc,$ we get
\begin{equation}
\begin{aligned}
\sum_{\nu=1}^{N}\left(1-\frac{c_{\nu}}{c}\right)\delta_{\kappa\nu}v^t_{-\nu}z^{\kappa}+\sum_{\kappa=1}^N\sum_{\nu=1}^N\frac{c_{\nu}}{c}v^t_{-\nu}\mathfrak{C}_{\kappa}\big(\mathfrak{P}_\mathcal{D}(\frac{\phi^{-\nu}_-}{L_-})\big)z^{\kappa}\\
=L_+^{-1}(z_{\mathtt{P}})\sum_{\kappa=1}^N\mathfrak{C}_{\kappa}\big(\mathfrak{P}_\mathcal{D}(\dfrac{v^i_-}{L_-})\big)z^\kappa, \quad
|z|<R_-.
\end{aligned}
\end{equation}
Above {equation} can be written in {a} symbolic manner as
\begin{equation}
a_{\kappa\nu}\chi_\nu=b_\kappa\quad\quad(\kappa, \nu=1, \dotsc, N),
\label{eqnaChi}
\end{equation}
where
\begin{equation}
\begin{aligned}
a_{\kappa\nu}&=\left(1-\frac{c_{\nu}}{c}\right)\delta_{\kappa\nu}+\frac{c_{\nu}}{c}\mathfrak{C}_{\kappa}(\mathfrak{P}_\mathcal{D}
\big(\frac{\phi^{-\nu}_-}{L_-}\big)),\\
\chi_\nu&=v^t_{-\nu},\quad
b_\kappa=L_+^{-1}(z_{\mathtt{P}})\mathfrak{C}_{\kappa}\big(\mathfrak{P}_\mathcal{D}(\dfrac{v^i_-}{L_-})\big).
\end{aligned}
\end{equation}
Formally, {applying the inversion of coefficient matrix in \eqref{eqnaChi}, i.e.,} ${\boldsymbol\chi}=\mathbf{A}^{-1}\mathbf{b}${,} gives $\{v^t_{x}\}_{x\in\mathcal{D}}$; substitution of this expression back in \eqref{vmsol}, via \eqref{defPt}, as well as \eqref{vpmsol} leads to the complete solution of the Wiener--Hopf equation.
Let $\tilde{a}_{\nu\kappa}$ denote the components of the inverse of $\mathbf{A}$. Then
\begin{equation}
v^t_{x}=\tilde{a}_{-x\kappa}L_+^{-1}(z_{\mathtt{P}})\mathfrak{C}_{\kappa}\big(\mathfrak{P}_\mathcal{D}(\dfrac{v^i_-}{L_-})\big).
\label{exactvmsol}
\end{equation}
The expression \eqref{exactvmsol} has been verified using a numerical solution (based on scheme described in Appendix D of \cite{sharma2015diffraction}) of the discrete Helmholtz equation \eqref{eq:main_equation} and assumed conditions on the crack faces for several choices of the damaged links; we omit the graphical plots of the comparison 
{as they are indistinguishable on the considered graph scale}.

Remark 1: When $\omega_2$ ($=\Im \omega$) is positive, it follows from the Krein conditions that there exists a unique solution in square summable sequences since only a finite number $N$ of damaged links are present. This is a statement on the lines of that {provided by Sharma} for the sharp crack-tip \cite{sharma2015diffraction,sharma2015diffraction3} and {the} rigorous results of {Ando} \cite{Ando}. The limiting case as $N\to\infty$ can be a different story altogether and it is not pursued here.

\section{{Examples of specific damage zones. }}
\label{specialdamage}

Choosing different values of the coefficients $c_{-j}$, $j\in[1,N]$ one can consider various damage zones.
Some of them are discussed below.

\subsection{{Completely destroyed zone.}}
Consider the simplest case when $c_{-x}\equiv 0$.
{(In fact, it is a bad choice of the left end of the cohesive zone).}
Then
\eqref{eqnNbyN} reduces to (using \eqref{sum_1_plus})
\begin{equation}
\sum_{x\in\mathcal{D}}v^t_xz^{-x}=L_+^{-1}(z_{\mathtt{P}})\mathfrak{P}_\mathcal{D}(\frac{v^i_-}{L_-})(z),\quad
 |z|<R_-,
\end{equation}
but $\sum_{x\in\mathcal{D}}v^t_xz^{-x}=\mathfrak{P}_\mathcal{D}(v_-+v^{i}_{-})(z)$, so that it is a special case of the complete exact solution given in \cite{sharma2015diffraction}, i.e., $v_-(z)=(L_+^{-1}(z_{\mathtt{P}})L_-^{-1}(z)-1)v^i_-(z)$, $|z|<R_-$ (see Eq. (2.29) in \cite{sharma2015diffraction} and Eq. (4.1b) in \cite{sharma2015diffraction3}). The detailed analysis and expressions of the solution based on latter appears in \cite{sharma2015diffraction3}
{ where the single crack was considered. It is natural, as this special case corresponds to a single (a bit longer) crack.}

\subsection{{"Healthy" (no damage) zone.}}
For the case $c_{-x}\equiv c$, the above extra equation \eqref{eqnNbyN} arises 
{again} due to a ``bad choice" of the origin 
{(cohesive crack-tip)} to define the half-Fourier transforms!
Consider the simplest case when $c_{-x}\equiv c$. Evidently, this case coincides with the previous 
{one} when $c_{-x}\equiv 0$, except for a shift in the origin from $(0, 0)$ to $(-N, 0)$
{(a single a bit shorter crack)}. Then
\eqref{eqnNbyN} reduces to (using \eqref{sum_1_plus})
\begin{equation}
\sum_{x\in\mathcal{D}}v^t_x\mathfrak{P}_\mathcal{D}L_-^{-1}\sum_{m=0}^{-x-1}\overline{l}_{+m}z^{-m}z^{-x}
=L_+^{-1}(z_{\mathtt{P}})\mathfrak{P}_\mathcal{D}L_-^{-1}v^i_-, \quad |z|<R_-.
\end{equation}
With {the substitution} $z\mapsto z^{-1}, x\mapsto-x$ {in above equation}, {we get}
\begin{equation}
\sum_{x=1}^{N}v^t_{-x}\mathfrak{P}_\mathcal{D}L^{-1}_+(z)\sum_{m=0}^{x-1}\overline{l}_{+m}z^{m}z^{-x}
=L_+^{-1}(z_{\mathtt{P}})\mathfrak{P}_\mathcal{D}L^{-1}_+(z)v^i_-(z^{-1}), \quad |z|>R_-^{-1},
\end{equation}
i.e.,
\begin{equation}
\sum_{x=1}^{N}v^t_{-x}\sum_{m=0}^{x-1}\overline{l}_{+m}z^{m}z^{-x}
=L_+^{-1}(z_{\mathtt{P}})v^i_-(z^{-1}), \quad |z|>R_-^{-1},
\end{equation}
{and finally,}
\begin{equation}
\sum_{x=1}^{N}v^t_{-x}z^{-x}L_+^{-1}(z)
=L_+^{-1}(z_{\mathtt{P}})v^i_-(z^{-1}), \quad |z|>R_-^{-1},
\end{equation}
Here the reference expression from \cite{sharma2015diffraction} and \cite{sharma2015diffraction3} is $v_+=(1-L_+^{-1}(z_{\mathtt{P}})L_+)v^i_-,  |z|>R_+${,} to which it agrees.

\subsection{A zone with continuously distributed damage.}

{Lets consider a relatively general case that {models a} real damage accumulation in the damage zone. In this case, one can {reasonably}
assume that at the crack-tip the stiffness of the interfacial zone is minimal (the damage is most pronounced) then increasing monotonically
and, finally at the other end of the zone, it takes the same magnitude as of non-damaged lattice. {A} typical representative of such interface is {the} exponential distribution
\[
c_{-x}=c\exp(\alpha x/N), \quad x\in\mathcal{D}.
\]
The parameter $\alpha$ regulates the rate of damage accumulation.
Note that $\alpha\gg1$ and $\alpha\ll1$ correspond to part (a) and part (b) of this section, respectively.
On Fig.~\ref{fig2} we present an illustration of $v^t_{x}$ given by
\eqref{exactvmsol}
for $N=100$. It is emphasized here that the graphical results for the {same choice can be} obtained using {the} numerical scheme (described in Appendix D of \cite{sharma2015diffraction}) {and these are found to} coincide with the plot in Fig.~\ref{fig2}(b).
\begin{figure}[!ht]
\center{\includegraphics[width=.7\linewidth]{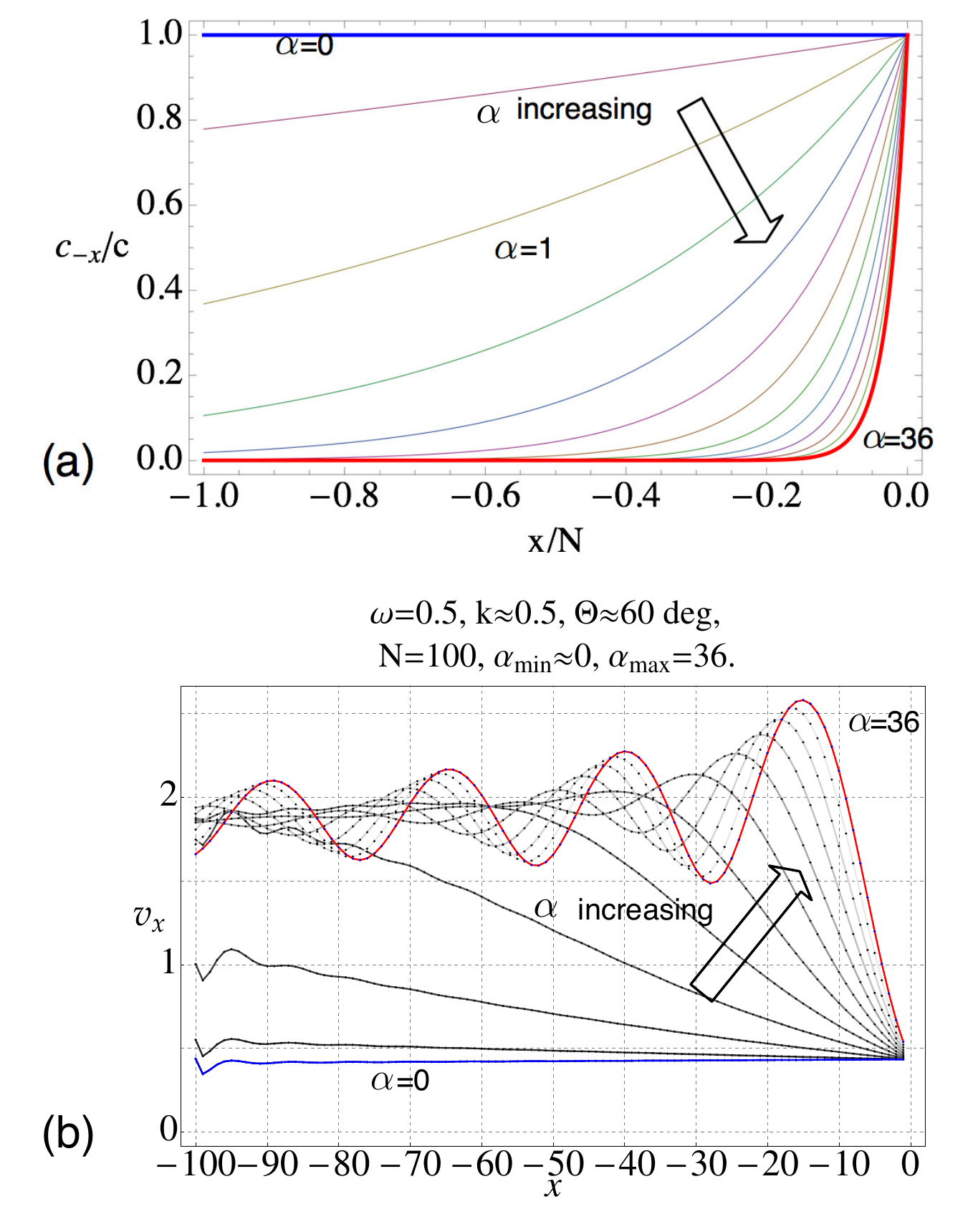}}
\caption[ ]{(a) Illustration of $c_{-x}$ with $c_{-x}=c\exp(\alpha x/N), x\in\mathcal{D}$. (b) Illustration of (total) $v_{x}$ given by \eqref{invfourier_v} for $N=100$. The curves in blue and red correspond to min and max value of $\alpha$, respectively.}
\label{fig2}
\end{figure}

{As one can see, {the presence of a} high gradient in the elastic properties of the cohesive zone significantly amplifies the local scattered field near the tip of the zone.
As a result, a pronounced damage should be expected exactly here that is consistent with the assumptions. However,
when $\alpha$ is close enough to zero, an opposite phenomenon happens as now the gradient is small while the jump of the material properties undergoes its maximum value (in fact it is equivalent to the second case above). It is thus important to compare which part of the damage zone can be subjected to higher risk for further damage. It is also evident that the angle of the incident wave $\theta$ may influence essentially the discussed effect.
Respective graphical results for the ratio $v_{-1}/v_{-N}$ are presented in Fig.\ref{fig_new1} showing impact of the incident wave frequency by considering two different normalised values $\omega=0.6$ and $\omega=1.2$. As expected large and small values of the parameter $\alpha$ determining the damage gradient inside the zone change the effect significantly. Namely for small value of $\alpha$ the left hand end of the damage zone is impacted by higher amplitudes and vice versa. In the right hand end of the zone (contacting with undamaged part of the zone) the effect is less straightforward. Also for the incident waves parallel to the crack ($\theta=0$ and $\theta=\pi$) the results are different. The first type can be in fact interpreted as the so-called feeding waves (see for example \cite{MMS_2009}) for the dynamic case.  }

\begin{figure}[!ht]
\center{\includegraphics[width=0.7\linewidth]{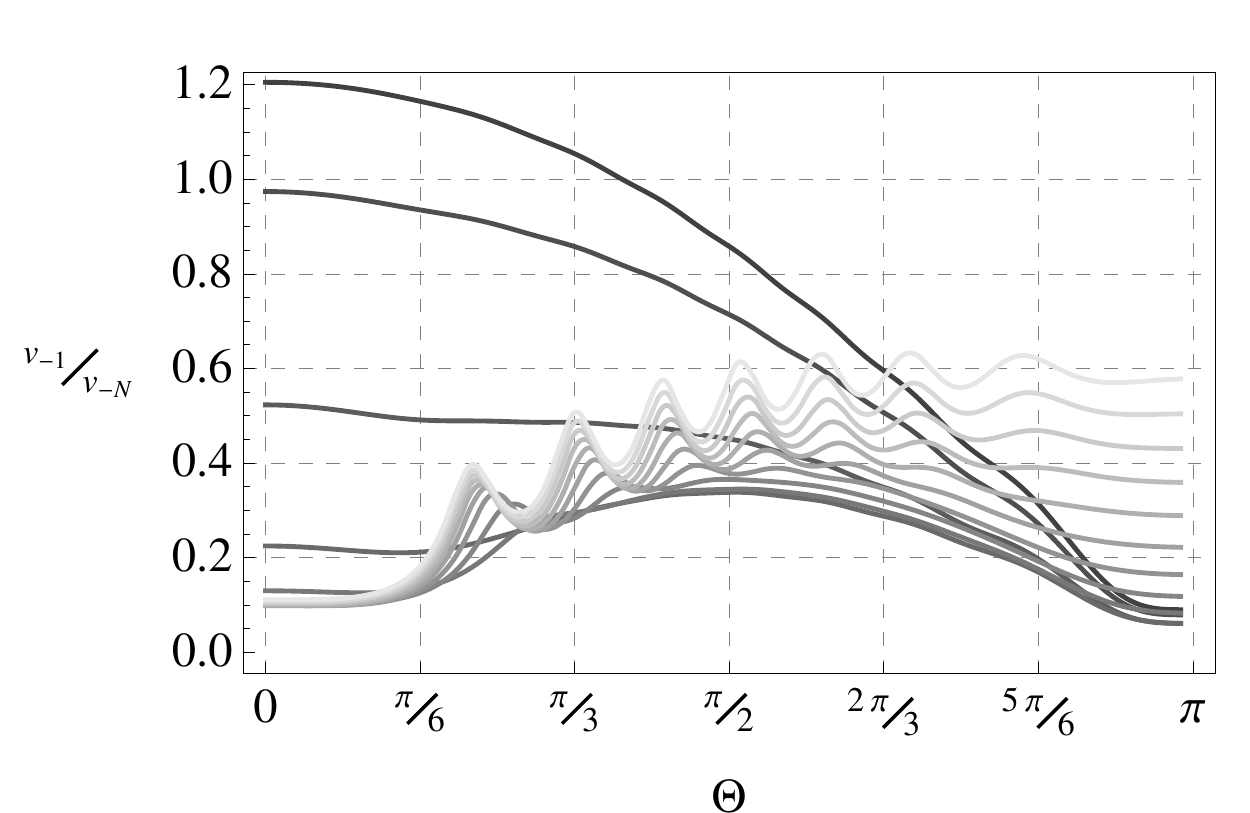}}
\center{\includegraphics[width=0.7\linewidth]{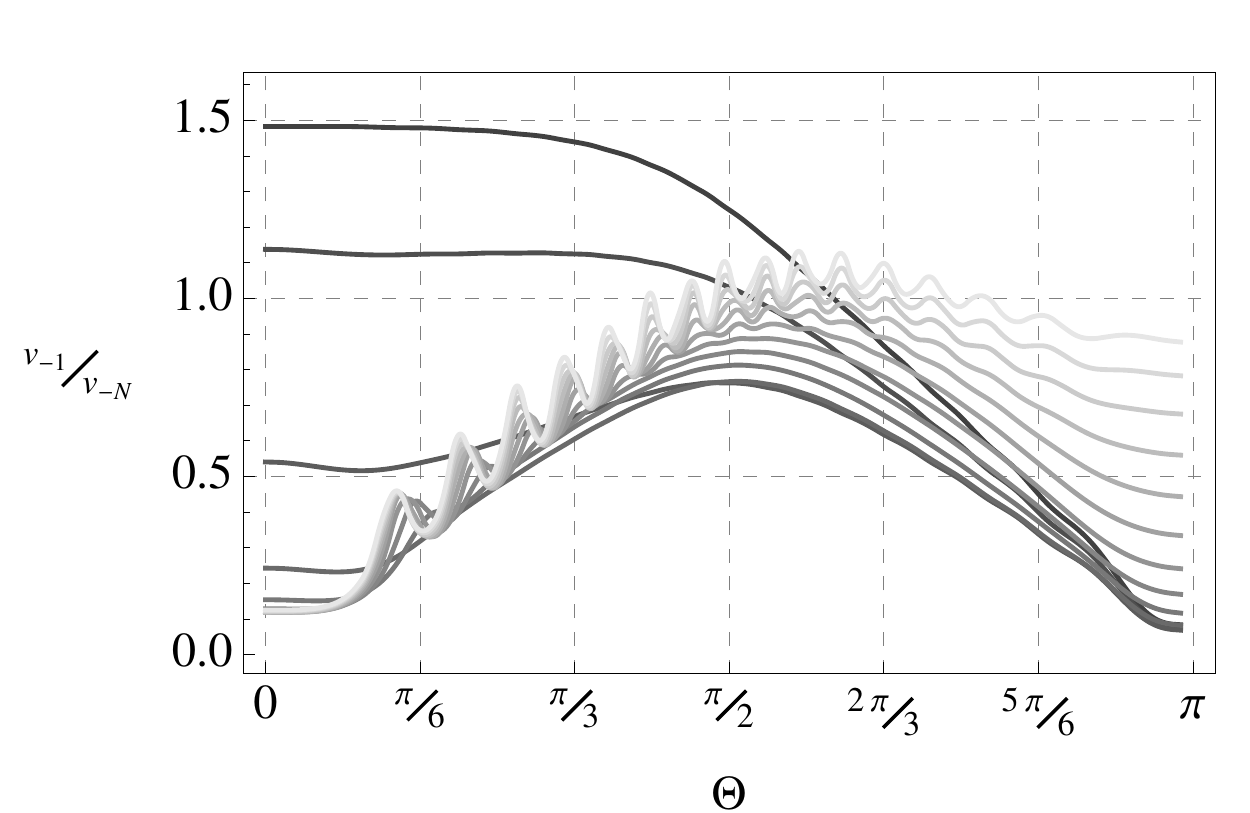}}
\caption[ ]{Ratio of the amplitudes at the ends of the damage zone of the length $N=40$. Top graph corresponds to $\omega=0.6$. The bottom one represents other normalised frequency $\omega=1.2$. Plots for a range of the parameter $\alpha=\{10^{-6}, 0.25, 1, 2.25, 4, 6.25, 9, 12.25, 16, 20.25, 25, 30.3, 36\}$) with darker shade for smaller $\alpha$.}
\label{fig_new1}
\end{figure}

{In Fig.\ref{fig_new2} we show in more detail influence of the big and small values of the parameter $\alpha$. Exact values of the parameters are depicted in the captions of the respective figures.}

\begin{figure}[!ht]
\center{\includegraphics[width=0.7\linewidth]{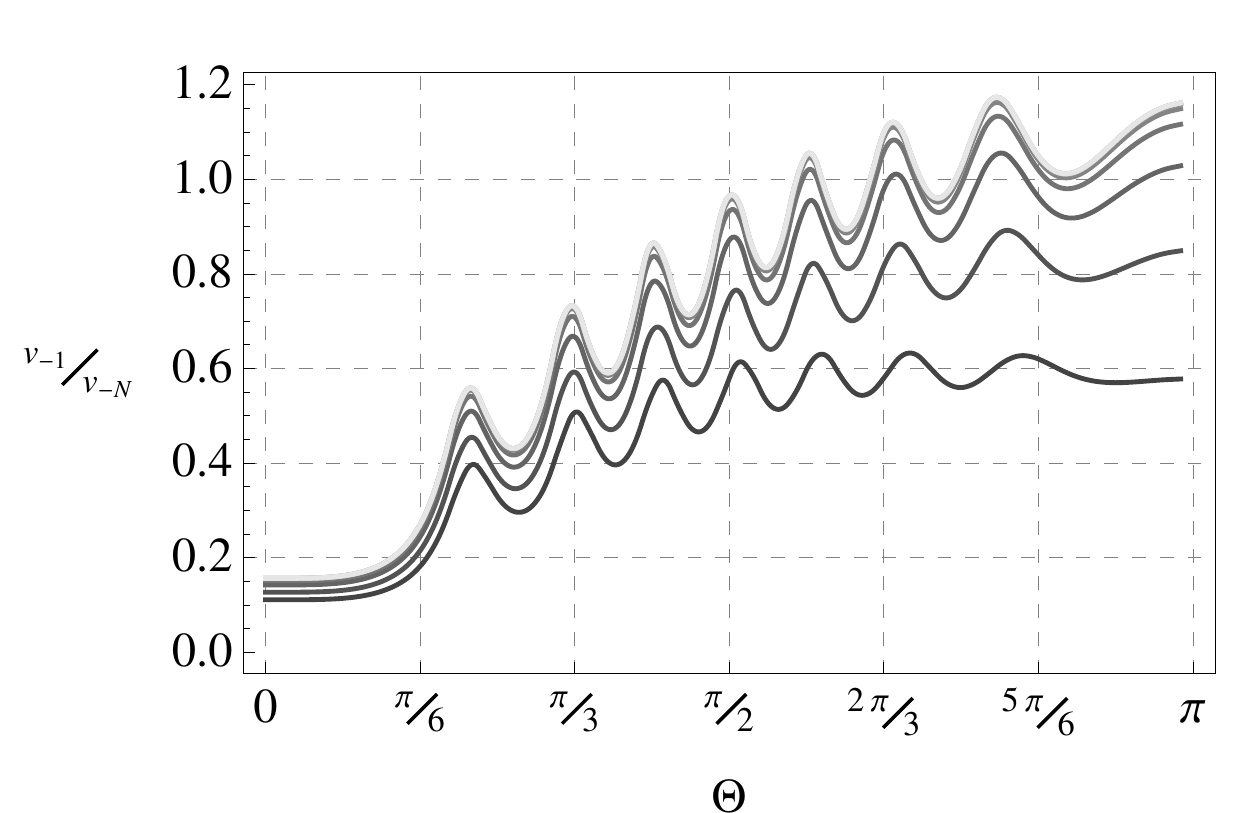}}
\center{\includegraphics[width=0.7\linewidth]{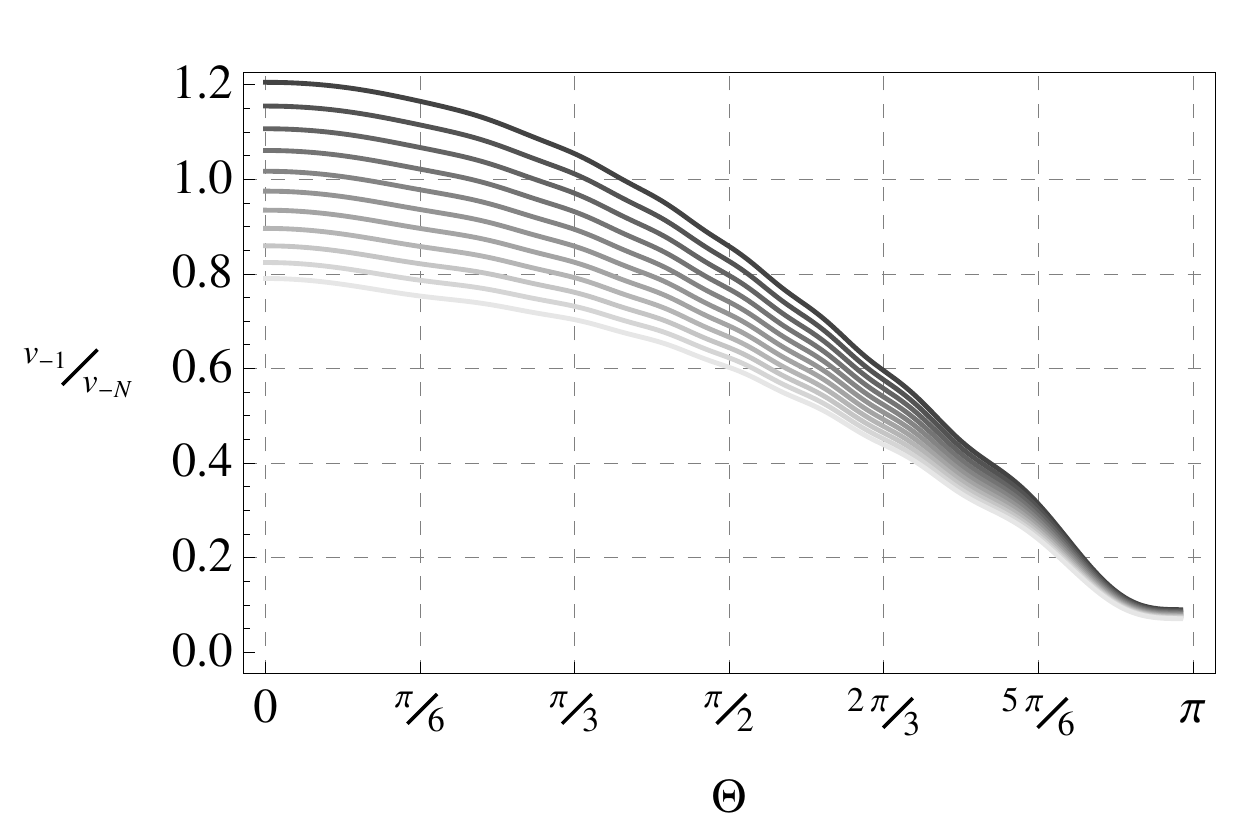}}
\caption[ ]{Ratio of the amplitudes at the ends of the damage zone of the length $N=40$ and the frequency $\omega=0.6$.
Top graph corresponds large values of $\alpha=\{36, 64, 100, 144, 196, 256, 324, 400, 484, 576, 676\}$), while the bottom one is for small values $\alpha=\{10^{-6}, 0.05, 0.1, 0.15, 0.2, 0.25, 0.3, 0.35, 0.4, 0.45, 0.5\}$). Plots for a range of $\alpha$ with darker shade for smaller $\alpha$.}
\label{fig_new2}
\end{figure}

\subsection{Damage represented by a bridge crack.}

Let $N$ be even. In the following, we will use the standard notation:
\begin{equation}
\begin{aligned}
\mathbb{Z}^+=\{0,1,2,\ldots\}, \quad \mathbb{Z}^-=\{-1,-2,\ldots\}, \\
\mathbb{Z}_e=\{0,\pm2,\pm4,\ldots\}, \quad \mathbb{Z}_o=\{\pm1,\pm3,\ldots\},\\
\mathbb{Z}^-_S=\mathbb{Z}_-\setminus\mathcal{D}, \quad\mathbb{Z}^+_S=\mathbb{Z}_+\cup\mathcal{D},
\end{aligned}
\label{eq:different_Z2}
\end{equation}
for different subsets of the set of entire numbers.
\begin{figure}[!ht]
\center{\includegraphics[width=.7\linewidth]{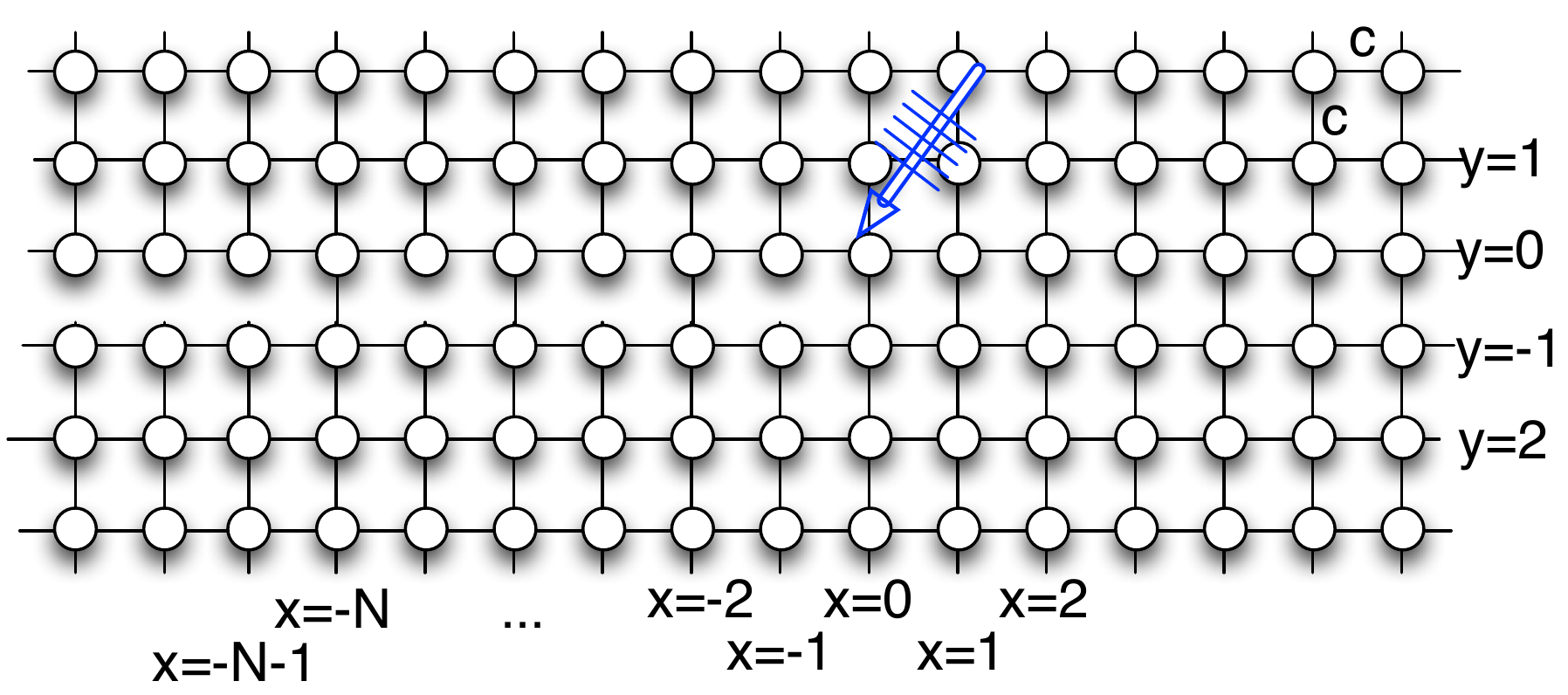}}
\caption[ ]{Geometry of the square lattice structure with partially open crack-tip and $N=2M=6.$}
\label{fig3}
\end{figure}
Consider the case when 
\[ 
c_{-x}=c, \quad x\in\mathcal{D}\cap\mathbb{Z}_e, \quad \mbox{and} \quad c_{-x}=0, \quad x\in\mathcal{D}\cap\mathbb{Z}_o
\] 
(see Fig. \ref{fig3}). Here $\max|\mathcal{D}\cap\mathbb{Z}_e|$ is $N$ which is replaced by $2M$ for convenience; thus the intact bonds on even sites in the cracked row begin at $x=-2M.$
The difference of \eqref{eq:sy=0b} and \eqref{eq:sy=-1b} becomes
\begin{equation}
\begin{aligned}
c(2u^s_{x,1}+v^s_{x+1,0}+v^s_{x-1,0}+(-5+\omega^2)v^s_{x,0})\\+2\Big(c-cH(x+2M)\delta_{x,e}-cH(x)\delta_{x,o}\Big)v^s_{x}\\
=-2\Big(c-cH(x+2M)\delta_{x,e}-cH(x)\delta_{x,o}\Big)v^i_{x}.
\end{aligned}
\label{eqvspec}
\end{equation}
Utilising the Fourier transform \eqref{fourier} to the equation \eqref{eqvspec} and taking into account the following representations of the functions
$u^F(z)=(u^s_x)^F$, $v^F(z)=(v^s_x)^F$
\begin{equation}
u^F(z)\equiv u^+(z)+u^-(z),\quad u^\pm(z)=\sum_{\mathbb{Z}^\pm}z^{-x}u_x,\quad
z\in\mathcal{A},
\label{fourier_pm}
\end{equation}
\begin{equation}
v^F(z)=v^+(z)+v^-(z), \quad v^\pm(z)=z^{2M}v^\pm_e(z)+v^\pm_o(z),
\label{eq:def_F_1}
\end{equation}
\begin{equation}
v^\pm_{e}(z)=z^{-2M}\sum_{\mathbb{Z}_S^\pm\cap\mathbb{Z}_{e}}z^{-x}v_x=\sum_{\mathbb{Z}^\pm\cap\mathbb{Z}_{e}}z^{-x}v_x,\quad
z\in\mathcal{A},
\label{eq:def_F_2}
\end{equation}
\begin{equation}
v^\pm_{o}(z)=\sum_{\mathbb{Z}^\pm\cap\mathbb{Z}_{o}}z^{-x}v_x,\quad
z\in\mathcal{A},
\label{eq:def_F_3}
\end{equation}
we get with
\begin{equation}
d^{-1}(z)=\lambda(z)-1-Q(z)=-(\lambda^{-1}+1),
\label{d(z)}
\end{equation}
\begin{equation}
d(z)^{-1}(z^{2M}v_e^++v_o^++z^{2M}v_e^-+v_o^-)+2 (z^{2M}v_e^-+v_o^-)=-2 (z^{2M}v_e^{i-}+v_o^{i-}),
\label{W-H_0}
\end{equation}
where we have taken into account that $z\mapsto-z$, since $v_e^+(z)=v_e^+(-z)$ while $v_o^+(z)=-v_o^+(-z)$.
Thus, we obtain {the} matrix Wiener--Hopf equation
\begin{equation}
\mathbf{A}(z)\mathbf{v}^+(z)+\mathbf{B}(z)\mathbf{v}^-(z)=\mathbf{f}(z),\quad
z\in\mathcal{A},
\label{eq:WH_0}
\end{equation}
where we have defined new plus and minus vector functions $\mathbf{v}^\pm(z)=(v^\pm_e,v^\pm_o)^\top$.
The components of the matrices $\mathbf{A}(s)$, $\mathbf{B}(s)$ and the right-hand side of the equation \eqref{eq:WH_0} are
\begin{equation}
\begin{matrix}
a_{11}=1,\quad b_{11}=(1+2d(z)),\quad a_{12}(z)=z^{-2M},\quad
\quad b_{12}(z)=z^{-2M}(1+2d(z)),\\[3mm]
a_{21}=z^{2M}, \quad b_{21}=z^{2M}(1+2d(-z)), \quad a_{22}(z)=-1, \quad b_{22}(z)=-(1+2d(-z)),
\end{matrix}
\label{eq:components}
\end{equation}
and
\begin{equation}
f_{1}(z)=-2(v_e^{i-}-z^{-2M}v_o^{i-})d(z),\quad
f_{2}(z)=-2(z^{2M}v_e^{i-}+v_o^{i-})d(-z).
\label{eq:componentsg}
\end{equation}
where $d(z)$ has been defined already above in \eqref{d(z)}.
Note $\mathbf{B}=\mathbf{A}+\mathbf{D}$, $d_{11}=2d(z)$, $d_{12}=z^{-2M}2d(z)$, $d_{21}=z^{2M}2d(-z)$, $d_{22}=-2d(-z)$.
Let
\begin{equation}
\begin{aligned}
\mathbf{C}&=\mathbf{I}+\mathbf{A}^{-1}\mathbf{D}=\mathbf{I}+\frac{1}{\det \mathbf{A}}\begin{pmatrix}
a_{22} & -a_{12}\\
-a_{21}& a_{11}
\end{pmatrix}\begin{pmatrix}
d_{11} & d_{12}\\
d_{21}& d_{22}
\end{pmatrix}\\
&=\mathbf{I}-
\begin{pmatrix}
d(z) & 0\\
0& -d(-z)
\end{pmatrix}
\begin{pmatrix}
1 & z^{-2M}\\
-z^{2M}& 1
\end{pmatrix}=\begin{pmatrix}
1-d(z) & -d(z)z^{-2M}\\
-d(-z)z^{2M}& 1+d(-z)
\end{pmatrix}.
\end{aligned}
\label{matrixWH}
\end{equation}
Equation \eqref{eq:WH_0} can be rewritten in an equivalent form:
\begin{equation}
\mathbf{v}^+(z)+\mathbf{C}(z)\mathbf{v}^-(z)=\mathbf{A}^{-1}(z)\mathbf{f}(z),\quad
z\in\mathcal{A},
\label{eq:WH_1}
\end{equation}
where
$\mathbf{C}(z)=\mathbf{A}^{-1}(z)\mathbf{B}(z).$
The matrix $\mathbf{C}$ possesses a structure which in general does not admit factorization by standard techniques for arbitrary $N$ (for $N=1$, perhaps).

{On the other hand, as it has been proven above, this special case can be reduced to the solution of $N$ linear algebraic equations (see also \cite{Sharmastaggerpair_exact}).} 
For example, {the} problem with a cohesive zone of similar geometry in \emph{continuous formulation} \cite{MishuV} cannot be reduced to a scalar Wiener-Hopf problem and {requires an application of} other numerical techniques \cite{mishuIII,mishuI,mishuII,Kisil}.}

\begin{figure}[!ht]
\center{\includegraphics[width=0.7\linewidth]{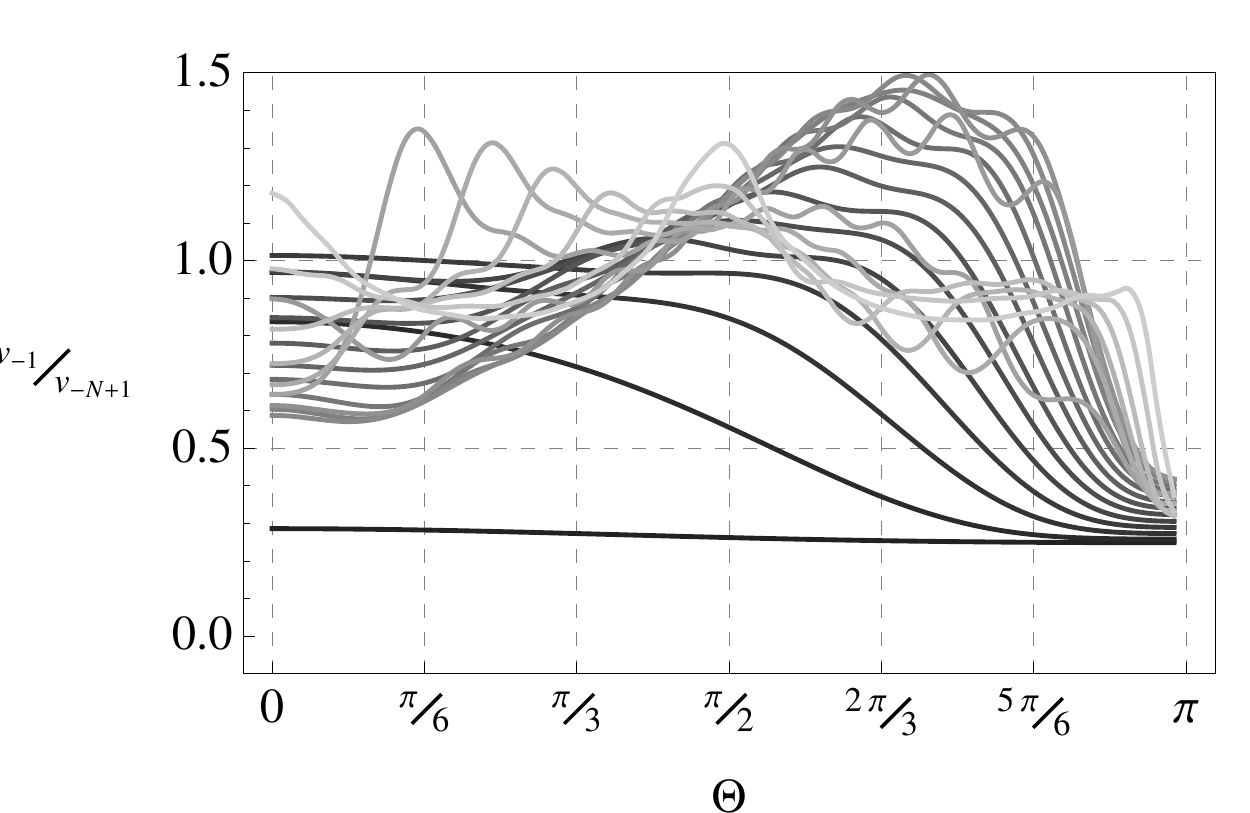}}
\center{\includegraphics[width=0.7\linewidth]{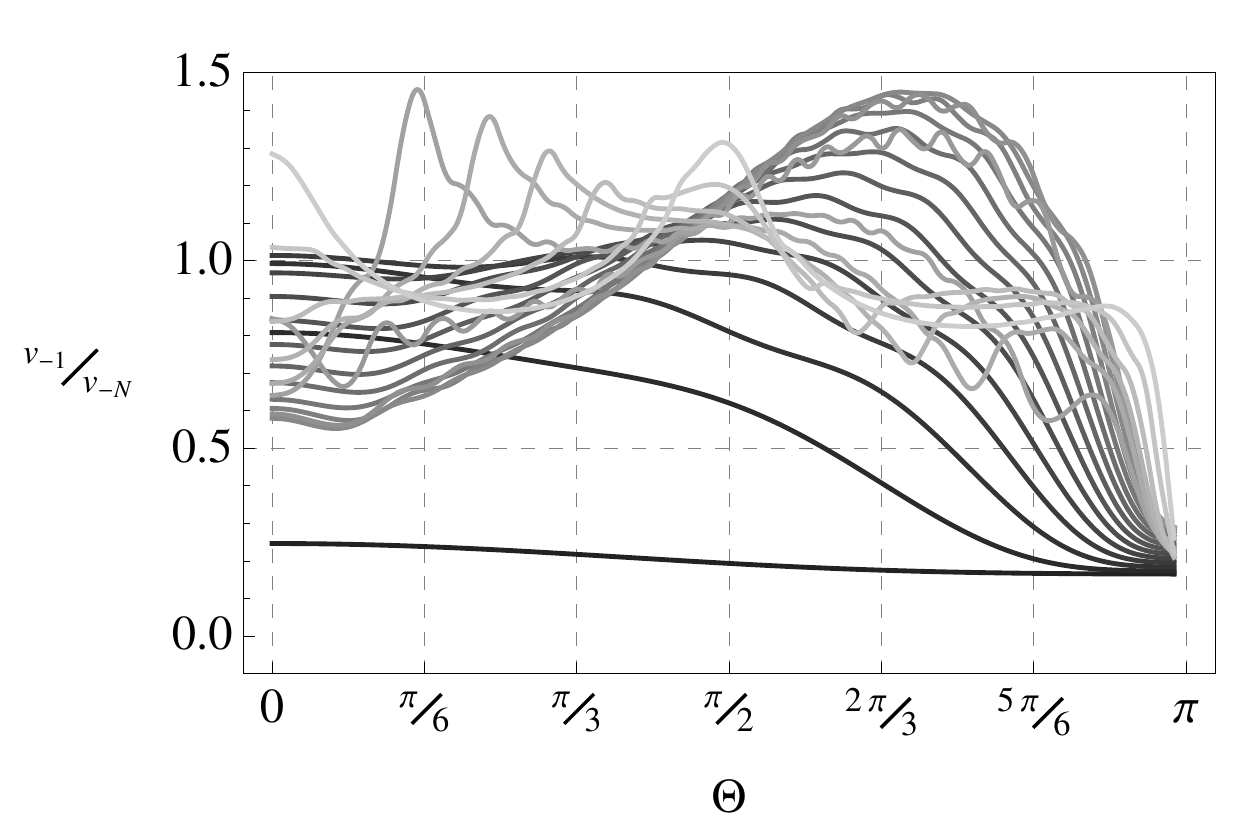}}
\caption[ ]{Length of the damage zone: $N=40$. Bridged bonds are $x=-1, -3, \dotsc, -N+1$ and bonds intact: $x=-2, -4, \dotsc, -N$. Plots for a range of frequencies of the incident waves: $\omega=\{0.01,$ $0.1,$ $0.2,$ $0.3,$ $0.4,$ 
$0.5,$ $0.6,$ $0.7,$ $0.8,$ $0.9,$ $1.0,$ $1.1,$ $1.2,$ $1.3,$ $1.4,$ $1.5,$ $1.6,$ $1.7,$ $1.8,$ $1.9,$ $1.95\}$, with darker shade used for smaller $\omega$.}
\label{fig_new3}
\end{figure}

{In Fig.\ref{fig_new3} we show the ratio of amplitudes in the two last points 
from the left-hand side of the damage zone to that on the right hand-side of the zone ($x=0$). 
Exact values of the parameters are depicted in the captions of the respective figures. Now we examine in more detail an impact of the frequency of the incident waves.}

In the context of the matrix kernel \eqref{matrixWH}, with the distinguished presence of the off-diagonal factors $z^{-2M}$ and $z^{2M}$, {the reduction to linear algebraic equation obtained} above, 
is reminiscent of that proposed for the Wiener--Hopf kernel with exponential phase factors that appear in several continuum scattering problems in fluid mechanics and fracture mechanics \cite{abPhase,abPlatesI,abPlatesII,abPlatesIII}, and their discrete analogues in the form of scattering due to a pair of staggered cracks and rigid constraints \cite{GMthesis,Gaurav2019_asymp,Sharmastaggerpair_exact}; both based on an exact solution of the corresponding staggerless case \cite{Heins1,Heins2,Abrahams0,Gaurav2019_pair}.

\section{Reconstruction of the scattered field}
\label{farfield}

Let $\mathtt{C}$ be a contour in the annulus $\mathcal{A}.$ By the inverse {Fourier} transform $u^s_{x,y}=\frac{1}{2\pi i}\int_{\mathtt{C}}u_{y}^F(z)z^{x-1}dz$, i.e.,
\begin{equation}
\begin{aligned}
u^s_{x,y}&=\frac{1}{2\pi i}\frac{1}{2}\int_{\mathtt{C}}v^F(z)\lambda^y(z) z^{x-1}dz,\quad x\in\mathbb{Z},\quad y=0,1,2,\ldots.
\label{invfourier_up}
\end{aligned}
\end{equation}
where $v^F$ is given by \eqref{invfourier_up2}.
For $y=0,1,2,\ldots,$
$u^s_{x,-y-1}=-u^s_{x,y}, x\in\mathbb{Z},$ due to skew-symmetry.
The total wave field is given by
\begin{equation}
u_{x,y}=u^s_{x,y}+u^i_{x,y},\quad x\in\mathbb{Z},\quad y\in\mathbb{Z}.
\end{equation}
Concerning the effect of the damage, using the decomposition $v^F(z)=v^F_a(z)+v_{\mathrm{P}}^F(z)$ \eqref{invfourier_up2}, it is easy to see that $v^F_a(z)$ coincides with the solution given in \cite{sharma2015diffraction}, i.e., it describes the scattering due to undamaged crack-tip; thus, the effect of the damage zone is represented by the second term $v_{\mathrm{P}}^F(z)$ in \eqref{invfourier_up2}.

The perturbation in the scattered field \eqref{invfourier_up} induced by the damage zone is given by
\begin{eqnarray}
\hat{u}_{x,y}&=&\frac{1}{2\pi i}\frac{1}{2}\int_{\mathtt{C}}v_{\mathrm{P}}^F(z)\lambda^y(z) z^{x-1}dz,\quad x\in\mathbb{Z},\quad y=0,1,2,\ldots,\nonumber\\
&=&\frac{1}{2\pi i}\frac{1}{2}\int_{\mathtt{C}}(L_+\mathtt{C}^{\mathrm{P}}_++L_-^{-1}\mathtt{C}^{\mathrm{P}}_-)\lambda^y(z) z^{x-1}dz\nonumber\\
&=&\frac{1}{2\pi i}\frac{1}{2}\int_{\mathtt{C}}(-\frac{1}{c}\sum_{m\in\mathcal{D}}c_{-m}v^t_m(L_+\phi^m_++L_-^{-1}\phi^m_-)+\frac{1}{c}\sum_{m\in\mathcal{D}}c_{-m}v^t_mz^{-m})\lambda^y(z) z^{x-1}dz\nonumber\\
&=&\frac{1}{2L_+(z_{\mathtt{P}})}\sum_{m\in\mathcal{D}}\frac{c_{-m}}{c}\tilde{a}_{-m\kappa}\mathfrak{C}_{\kappa}\big(\mathfrak{P}_\mathcal{D}\frac{v^i_-}{L_-}\big)\frac{1}{2\pi i}\int_{\mathtt{C}}\Lambda_m(z)\lambda^y(z) z^{x-1}dz,
\label{invfourier_pert_upfinal}
\end{eqnarray}
where
\begin{equation}
\Lambda_m(z)=z^{-m}-(L_+(z)\phi^m_+(z)+L_-^{-1}(z)\phi^m_-(z)), \quad m\in\mathcal{D}.
\label{invfourier_pert_up2}
\end{equation}

For $\xi\sqrt{x^2+y^2}\gg1$ and $\omega/c\in (0,2)$
where $\xi\sim\omega/c$ is related to the wave number of incident wave, a far-field approximation of the exact solution \eqref{invfourier_up} can be constructed; also {an} analogous result holds for $\omega/c\in (2,2\sqrt{2})$.
It is sufficient for our purpose to focus on the effect of the damage zone $\mathcal{D}$ so that we investigate the far-field approximation of \eqref{invfourier_pert_upfinal}, i.e., mainly associated with the expression of $\Lambda_m$ given by \eqref{invfourier_pert_up2} for each $m\in\mathcal{D}$.
Following \cite{sharma2015diffraction}, the approximation of far-field can be obtained using the stationary phase method \cite{felsen}.
The substitution $z=e^{-i\xi}$ maps the contour $\mathtt{C}$ into a contour $C_{\xi}$. In terms of polar coordinates {$(R, \theta)$}, {the lattice point $(x,y)$ can be expressed as}
\begin{equation}
x=R\cos\theta, y=R\sin\theta.
\end{equation}
{Let}
\begin{equation}
\Phi(\xi)=\eta(\xi)\sin\theta-\xi\cos\theta, \eta(\xi)=-i\log{\lambda}(e^{-i\xi}),
\label{phi}
\end{equation}
The function $\Phi$ \eqref{phi} possesses a saddle point \cite{Erdelyi1,Felsen1} at $\xi=\xi_{{S}}$ on $C_{\xi}$, with $\Phi'(\xi_{{S}})=\eta'(\xi_{{S}})\sin\theta-\cos\theta=0, \Phi''(\xi_{{S}})=\eta''(\xi_{{S}})\sin\theta\ne 0,$ which is same as that discussed in \cite{sharma2015diffraction}.
Omitting the details of the calculations, it is found that
\begin{equation}
\begin{aligned}
\hat{u}_{x,y}{\sim}
&\frac{1}{2\sqrt{\pi}}\frac{1+i\text{ sign}(\eta''(\xi_{S}))}{2c}\frac{\lambda^y(e^{-i\xi_{S}}) e^{-i\xi_{S}(x-1)}}{(R|\eta''(\xi_{S})|\sin\theta)^{1/2}}\sum_{m\in\mathcal{D}}(c_{-m}v^t_m\Lambda_m(e^{-i\xi_{S}})),
\end{aligned}
\label{uhatfarfield}
\end{equation}
as $\omega R/c\to\infty$. The expression \eqref{uhatfarfield} has been verified using a numerical solution of the discrete Helmholtz equation (based on scheme described in Appendix D of \cite{sharma2015diffraction}); a graphical demonstration of the same is omitted in the paper.

\section{Concluding remarks}
\label{concl}

We have shown how the scattering problem in square lattice with an infinite crack having a {damage} zone near the crack-tip of arbitrary properties can be effectively solved by We were able to reduce it to a scalar Wiener-Hopf method. We have used a new method that utilises specific discrete properties of the system under consideration. It consists of solving an auxiliary $N \times N$ system of linear equations with a unique solution (Remark 1).
Effectiveness of the method has been highlighted by some numerical examples and the constructed asymptotic expression of the scattered field at infinity. Analysis of the solution near two ends of the damage zone and at infinity can be used in
{a} {non-destructive} testing procedure{,} among other applications. The method may be useful to solve other {matrix} Wiener--Hopf problems appearing {in analysis of the} dynamics of discrete structures with defects.
{the} Wiener--Hopf technique. {Indeed, the discrete scattering problem for the bridge damage zone has been written in a vectorial problem with $2\times2$ matrix-kernel and simultaneously transformed it, by the aforementioned approach, to a scalar one (modulo the accompanying linear algebraic equation). This gives rise for a hope for building a close form standard procedure that allows for effective factorisation of similar matrices of an arbitrary size.}

\section*{Acknowledgement}
BLS acknowledges the partial support of SERB MATRICS grant MTR/2017/000013.
GM acknowledges financial support from the ERC Advanced Grant
"Instabilities and nonlocal multiscale modelling of materials",  ERC-2013-ADG-340561-INSTABILITIES.
He is also thankful to the Royal Society for the Wolfson Research Merit  Award and the
Isaac Newton Institute for Mathematical Sciences for {the} Simon's Fellowship.

The authors would like to thank the Isaac Newton Institute for
Mathematical Sciences, Cambridge, for support and hospitality during the
programme "Bringing pure and applied analysis together via the Wiener-Hopf technique, its generalisations and applications" where work on this paper was
completed. This programme was supported by EPSRC grant no EP/R014604/1.

\section*{Authors' Contributions}
{Both authors contributed equally to the writing of the manuscript, the problem formulation as well as the discussion/interpretation of the results in the context of special cases of the damage zone.
The Wiener--Hopf analysis and the technique of reduction to a linear algebraic equation as well as the asymptotic approximation of the scattered field is due to BLS, in addition to the numerical calculations based on the semi-analytical approach and a direct numerical solution of the scattering problem.
The formulation of bridged crack and the relationship of the scattering problem with 2x2 Matrix kernels is due to GM.
}

\printbibliography

\end{document}